\documentclass{aastex63}
\usepackage{graphicx}
\usepackage{gensymb}

\usepackage{graphicx}
\usepackage{wrapfig}
\usepackage{sidecap}
\usepackage{float}
\usepackage{array}
\usepackage{booktabs}
\accepted{ApJ Letters, October 1, 2020}
\shorttitle{Investigating Callisto's 4-$\micron$ band}
\shortauthors{Cartwright et al.}

\begin{document}

\title{Evidence for sulfur-bearing species on Callisto's leading hemisphere: \\ Sourced from Jupiter's irregular satellites or Io?}

\correspondingauthor{Richard J. Cartwright}
\email{rcartwright@seti.org}

\author[0000-0002-6886-6009]{Richard J. Cartwright$^a$}
\affiliation{The Carl Sagan Center at the SETI Institute \\
189 Bernardo Avenue, Suite 200\\
Mountain View, CA 94043, USA}
\footnote{$^a$Visiting Astronomer at the Infrared Telescope Facility, which is operated by the University of Hawaii under contract NNH14CK55B with the National Aeronautics and Space Administration. }

\author{Tom A. Nordheim}
\affiliation{Jet Propulsion Laboratory \\
	4800 Oak Grove Drive\\
	Pasadena, CA 91109, USA}

\author{Dale P. Cruikshank}
\affiliation{NASA-Ames Research Center\\
	Mail Stop 245-1\\
	Building N245, Room 204\\
	P.O. Box 1\\
	Moffett Field, CA 94035, USA}

\author{Kevin P. Hand}
\affiliation{Jet Propulsion Laboratory \\
	4800 Oak Grove Drive\\
	Pasadena, CA 91109, USA}

\author{Joseph E. Roser}
\affiliation{The Carl Sagan Center at the SETI Institute \\
	189 Bernardo Avenue, Suite 200\\
	Mountain View, CA 94043, USA}
\affiliation{NASA-Ames Research Center\\
	Mail Stop 245-1\\
	Building N245, Room 204\\
	P.O. Box 1\\
	Moffett Field, CA 94035, USA}

\author{William  M. Grundy}
\affiliation{Lowell Observatory \\
	1400 W Mars Hill Road\\
	Flagstaff, AZ 86001, USA}
\affiliation{Northern Arizona University \\
	S San Francisco Street\\
	Flagstaff, AZ 86011, USA}

\author{Chloe B. Beddingfield}
\affiliation{The Carl Sagan Center at the SETI Institute \\
	189 Bernardo Avenue, Suite 200\\
	Mountain View, CA 94043, USA}
\affiliation{NASA-Ames Research Center\\
	Mail Stop 245-1\\
	Building N245, Room 204\\
	P.O. Box 1\\
	Moffett Field, CA 94035, USA}

\author{Joshua P. Emery}
\affiliation{Northern Arizona University \\
	S San Francisco Street\\
	Flagstaff, AZ 86011, USA}



\begin{abstract}
We investigated whether sulfur-bearing species are present on the icy Galilean moon Callisto by analyzing eight near-infrared reflectance spectra collected over a wide range of sub-observer longitudes. We measured the band areas and depths of a 4-$\micron$ feature in these spectra, which has been attributed to sulfur dioxide (SO$_2$), as well as carbonates, in previously collected datasets of this moon. All eight spectra we collected display the 4-$\micron$ band. The four spectra collected over Callisto's leading hemisphere display significantly stronger 4-$\micron$ bands compared to the four trailing hemisphere spectra ($>$3$\sigma$ difference). We compared the central wavelength position and shape of Callisto's 4-$\micron$ band to laboratory spectra of various sulfur-bearing species and carbonates. Our comparison demonstrates that Callisto's 4-$\micron$ band has a spectral signature similar to thermally-altered sulfur, as well as a 4.025 $\micron$ feature attributed to disulfanide (HS$_2$). Our analysis therefore supports the presence of S-bearing species on Callisto but is not consistent with the presence of SO$_2$. The significantly stronger 4-$\micron$ band detected on Callisto's leading hemisphere could result from collisions with H$_2$S-rich dust grains that originate on Jupiter's retrograde irregular satellites or implantation of magnetospheric S ions that originate from volcanic activity on Io. Alternatively, S-bearing species could be native to Callisto and are exposed by dust collisions and larger impacts that drive regolith overturn, primarily on its leading side. 
\end{abstract} 



\section{Introduction} 
Sulfur is one of the key elements required for life as we know it. Measuring the abundance, distribution, and spectral signature of sulfur on ocean worlds like the icy Galilean moons Callisto, Ganymede, and Europa is therefore important for assessing their astrobiological potential \citep[e.g.,][]{hendrix2019nasa}. The innermost Galilean moon Io erupts substantial quantities of S-bearing neutrals into orbit, which are ionized and subsequently trapped in Jupiter's magnetosphere \citep[e.g.,][]{schneider2007io}. These sulfur ions are delivered primarily to the trailing hemispheres of Europa and Ganymede by Jupiter's co-rotating plasma, spurring a cascade of radiolytic surface chemistry \citep[e.g.,][]{johnson2004radiation}. The flux of magnetospheric S ions is lower at the orbit of Callisto but is still sufficient to drive radiolytic modification of its surface. 

Analysis of spectra collected by the Near Infrared Mapping Spectrometer (NIMS) on the Galileo spacecraft has revealed the presence of an absorption band centered near 4.25 $\micron$, which likely results from carbon dioxide (CO$_2$) \citep[e.g.,][]{carlson1996near, mccord1997organics, mccord1998non}. This 4.25-$\micron$ CO$_2$  band is distributed across Callisto's trailing hemisphere, producing a `bullseye' shaped pattern centered around 270$\degree$ longitude (i.e., the center of its trailing hemisphere) \citep[e.g.,][]{hibbitts2000distributions} and is likely formed by Jupiter's co-rotating plasma flowing onto Callisto's trailing side. CO$_2$ is also present on Callisto's leading hemisphere, but displays a different distribution, with concentrated deposits of CO$_2$ spatially associated with impact craters and their ejecta blankets \citep{hibbitts2000distributions,hibbitts2002co2}. 

Other prominent absorption bands are present in NIMS spectra of Callisto, including a feature centered between 4.02 - 4.05 $\micron$ (hereafter referred to as the `4-$\micron$' band), which was originally attributed to sulfur dioxide (SO$_2$) on Europa and Ganymede, as well as on Callisto \citep[e.g.,][]{carlson1996near,mccord1997organics, mccord1998non,hibbitts2000distributions}. Analysis of NIMS spectra demonstrated that the 4-$\micron$ band is strongest on the trailing hemispheres of Europa and Ganymede, consistent with implantation of S ions from Io. In contrast, Callisto's 4-$\micron$ band appears to be stronger on its leading hemisphere, hinting at a possibly different origin from the 4-$\micron$ band detected on Europa and Ganymede \citep{hibbitts2000distributions}. 
The 4-$\micron$ band has a mottled distribution across Callisto's leading hemisphere, without clear association with impact features or other landforms, unlike CO$_2$ \citep[e.g.,][]{hibbitts2000distributions}, suggesting that the 4-$\micron$ feature could have a different origin from CO$_2$ on this moon. 

In contrast to the sharp shape and strong absorption exhibited by the 4.07-$\micron$ SO$_2$ frost combination band (\textit{$\nu$}$_1$ + \textit{$\nu$}$_3$) \citep[e.g.,][]{nash1995laboratory} detected on Io \citep[e.g.,][]{fanale1979significance,cruikshank1980infrared}, Callisto's 4-$\micron$ band is more subtle, has a rounded shape, and is shifted to shorter wavelengths ($\sim$4.02 - 4.05 $\micron$). These spectral differences have led some studies to cast doubt on the presence of S-bearing species like SO$_2$ on Callisto, suggesting instead that C-rich species like calcium carbonate (CaCO$_3$) or sodium carbonate (Na$_2$CO$_3$) might be better candidates to explain the 4-$\micron$ band \citep{johnson2004radiation}. Similarly, analysis of data collected by the Ultraviolet Spectrometer (UVS) onboard Galileo indicates that the 0.28-$\micron$ feature detected on Callisto's leading side is more consistent with the spectral slope of C-rich material \citep{hendrix2008callisto}, as opposed to an SO$_2$ band suggested by prior work \citep[e.g.,][]{lane1997iue,noll1997detection}. 

Therefore, the origin and composition of Callisto's 4-$\micron$ band, and the possible presence of sulfur on this moon, remain uncertain. Complicating matters, the often low signal-to-noise (S/N) and low resolving power (R $\sim$200) of NIMS spectra has made assessment of the spectral signature and distribution of Callisto's 4-$\micron$ band more challenging. To investigate the 4-$\micron$ band further, we collected new near-infrared (NIR) reflectance spectra of Callisto. We detected the 4-$\micron$ band in these new ground-based spectra, along with other absorption bands between 2.8 - 5.0 $\micron$. We used these data to characterize the spectral signature and longitudinal distribution of Callisto's 4-$\micron$ band, providing new constraints on the presence and origin of S-bearing species. 

\section{Data and Methods}

\subsection{Observations and Data Reduction} 
We observed Callisto on eight different nights in May and June, 2020 using the NIR SpeX spectrograph/imager \citep[e.g.,][]{rayner2003spex} on NASA's Infrared Telescope Facility (IRTF), operating in long cross-dispersed mode, spanning 1.98 - 5.3 $\micron$ over seven spectral orders (`LXD$\_$long' mode). All LXD$\_$long observations were made with a 0.3'' wide by 1'' long slit, providing an average resolving power of 2500 in each spectral order. Prior to each observation, the SpeX slit was oriented parallel to Callisto's poles, and then placed over the center of Callisto's disk, thereby ensuring that the sub-observer longitude was fully covered by the slit. These spectra were collected using `AB' nodding, where all targets are observed in two different positions, separated by 7.5''. Each `A' and `B' frame had a maximum exposure length of 3 s to prevent saturation at wavelengths $>$4.1 $\micron$. During data reduction, these `A' and `B' frames were separated into sequential pairs, and the `B' frames were subtracted from the `A' frames, thereby performing first order sky emission correction. 

Flatfield frames were generated with an internal quartz lamp to correct for pixel variations across the chip. To perform wavelength calibration of these data, we collected arc lamp frames, using an internal argon lamp. Because argon emission lines are weak in L/L' and M bands ($\sim$2.8 - 5.3 $\micron$), we utilized night sky emission lines to calibrate the longer wavelength data. Data extraction, flatfielding, wavelength calibration, and spectral order merging were conducted using the Spextool data reduction suite \citep{cushing2004spextool}, as well as custom programs. All extracted spectra were divided by standard star spectra, observed on the same night immediately before and after observations of Callisto (within $\pm 0.1$ airmass), to provide additional telluric correction and remove the solar spectrum. The standard stars we observed were HD 189893 (G0V) and SA 112 1333 (F8). After dividing each spectrum by the standard star, all Callisto frames collected on the same night were co-added to boost S/N. The observation details are summarized in Table 1. 

\subsection{Band Parameter Analyses} 
To assess the spectral signature and longitudinal distribution of the 4-$\micron$ band, we measured its center, area, and depth in each collected spectrum with a custom band parameter analysis program that our team has used previously to analyze reflectance data of icy moons \citep[e.g.,][]{cartwright2015distribution,cartwright2020probing,cartwright2020evidence}. After reading in the data, the program fit the continuum of each spectrum between 3.926 and 4.114 $\micron$ to measure the 4-$\micron$ band. Next, the program divided each 4-$\micron$ band by its continuum, and measured the area of the resulting continuum-divided band using the trapezoidal rule. The program then used Monte Carlo simulations to estimate the uncertainties for these band area measurements by resampling the 1$\sigma$ errors of each spectral channel within each band (iterated 20,000 times). To measure the depth of each spectrum's 4-$\micron$ band, the program identified the spectral channel closest to 4.02 $\micron$ as the band center, and then calculated the mean  reflectance using all spectral channels within $\pm 0.015$ $\micron$. To calculate the band depth, the program subtracted these mean reflectances from 1. To estimate measurement errors, the program added the 1$\sigma$ uncertainties of all spectral channels included in the mean reflectance measurement, in quadrature, and then divided by the number of channels (n) to calculate the mean uncertainty ($\overline{\sigma}$). To estimate the point-to-point variation in each band, the program calculated the standard deviation of the mean ($\sigma$$\overline{_x}$ = $\sigma$/$\sqrt{n}$). The program then calculated the final uncertainty for each band depth measurement by adding $\overline{\sigma}$ and $\sigma$$\overline{_x}$ in quadrature. An example of our band measurement procedure is shown in Figure A1. 

\setlength{\heavyrulewidth}{1.2pt}
\setlength{\abovetopsep}{-3pt} 
\setlength{\cmidrulekern}{-1pt}

\begin{table}[tbp]
	\centering
	 \caption {IRTF/SpeX observations of Callisto.}
		\hskip-0.8cm\begin{tabular}{*9c}
			 \toprule
			\begin{tabular}[c]{@{}l@{}}\hspace{-1 cm}Spectrum\\  \hspace{-1 cm}Number \end{tabular} &\begin{tabular}[c]{@{}l@{}}\hspace{-1 cm}Sub-Observer\\  \hspace{-1 cm}Longitude ($\degree$)\end{tabular} & \begin{tabular}[c]{@{}l@{}} \hspace{-1 cm}Sub-Observer \\  \hspace{-1 cm}Latitude  ($\degree$)\end{tabular} & \begin{tabular}[c]{@{}l@{}}\hspace{-1 cm}Angular\\  \hspace{-1 cm}Diameter($\arcsec$)\end{tabular} & \begin{tabular}[c]{@{}l@{}} \hspace{-1 cm}Phase \\  \hspace{-1 cm}Angle ($\degree$)\end{tabular} & UT Date & \begin{tabular}[c]{@{}l@{}} \hspace{-1 cm}UT Time  \\  \hspace{-1 cm}(mid-expos)\end{tabular} & \begin{tabular}[c]{@{}l@{}} \hspace{-1 cm}Integration \\ \hspace{-1 cm}Time   (min)\end{tabular} & \begin{tabular}[c]{@{}l@{}} \hspace{-1 cm}Airmass \\  \hspace{-1 cm}Range  \end{tabular}  \\
			\midrule
			1 & 20.7 & -1.0 &  1.56 & 5.78 & 20/06/15 & 13:00 & 24 &  1.33 - 1.39 \\
		    2 &	64.0 & -1.0 & 1.56 & 5.45 & 20/06/17 & 13:00 & 25 &  1.33 - 1.46 \\
			3 & 120.6 & -1.1 &  1.52 & 7.75 & 20/06/03 & 12:45 & 26 & 1.32 - 1.49 \\
			4 & 164.7 & -1.1 & 1.53 & 7.44 & 20/06/05 & 13:30 & 25 & 1.31 - 1.39 \\
			5 & 200.7 & -1.1 & 1.47 & 9.37 & 20/05/21 & 13:45 & 24 & 1.32 - 1.40 \\
			6 & 244.0 & -1.1 & 1.47 & 9.13 & 20/05/23 & 13:45 & 25 & 1.32 - 1.39 \\
			7 & 287.2 & -1.1 & 1.48 & 8.88 & 20/05/25 & 13:45 & 28 & 1.33 - 1.41 \\
			8 & 337.8 & -1.0 & 1.55 & 6.11& 20/06/13 & 13:15 & 26 & 1.33 - 1.42 \\
			\bottomrule
		\end{tabular} 
\end{table}

\section{Results and Analysis} 

\subsection{IRTF/SpeX Spectra of Callisto} 
All eight of the SpeX spectra we collected display evidence for the 4-$\micron$ band, centered near 4.02 $\micron$ (Figures 1, A2). Our 4-$\micron$ band area and depth measurements are reported in Table 2 (along with their 1$\sigma$ uncertainties). These measurements confirm the presence of the 4-$\micron$ band in all eight spectra, demonstrating that this feature is present across Callisto's surface. Furthermore, the four spectra collected over Callisto's leading hemisphere (1 - 180$\degree$ longitude) show significantly stronger 4-$\micron$ bands compared to the four spectra collected over its trailing hemisphere (181 - 360$\degree$ longitude), which is consistent with previous analyses \citep[e.g.,][]{hibbitts2000distributions} (Figure A3). We further investigate the spatial distribution of the 4-$\micron$ band in the following subsection 3.2.

The SpeX spectra we collected display other spectral features that we briefly describe here. All collected SpeX spectra display the strong 3-$\micron$ H$_2$O ice band, as well as a subtle 3.1 $\micron$ Fresnel peak \citep[e.g.,][]{mastrapa2009optical}. These H$_2$O ice features were detected previously in ground-based \citep[e.g.,][]{pollack1978near,calvin1993spectral} and NIMS \citep[e.g.,][]{carlson1996near} spectra. The SpeX data display a wide band centered near  4.57 $\micron$, hereafter referred to as the `4.6-$\micron$' band, which was originally detected by NIMS and attributed to CN-bearing species \citep[e.g.,][]{mccord1998non}, or alternatively, carbon suboxide (C$_3$O$_2$, \citealt{johnson2004radiation}). Similar to the 4-$\micron$ band, the SpeX spectra we collected indicate that the 4.6-$\micron$ band is stronger on Callisto's leading hemisphere. The SpeX spectra also display a band centered near 3.87 $\micron$, hereafter referred to as the `3.9-$\micron$' band. This band was previously detected by NIMS and attributed to  SH-bearing materials \citep[e.g.,][]{mccord1998non}, as well as carbonic acid (H$_2$CO$_3$, e.g., \citealt{hage1998carbonic}). The 3.9-$\micron$ band in our SpeX spectra does not appear to display strong leading/trailing asymmetries in its distribution, and it is notably weaker than the 3.9-$\micron$ band detected by NIMS. Hints of other, more subtle bands are present in these spectra as well, which we discuss in Appendix A4 (Figure A4). Although of great interest, quantitative analysis of these other absorption features is beyond the scope of this paper and will be included in future work. 

\begin{figure}[h]
	\includegraphics[scale=0.70]{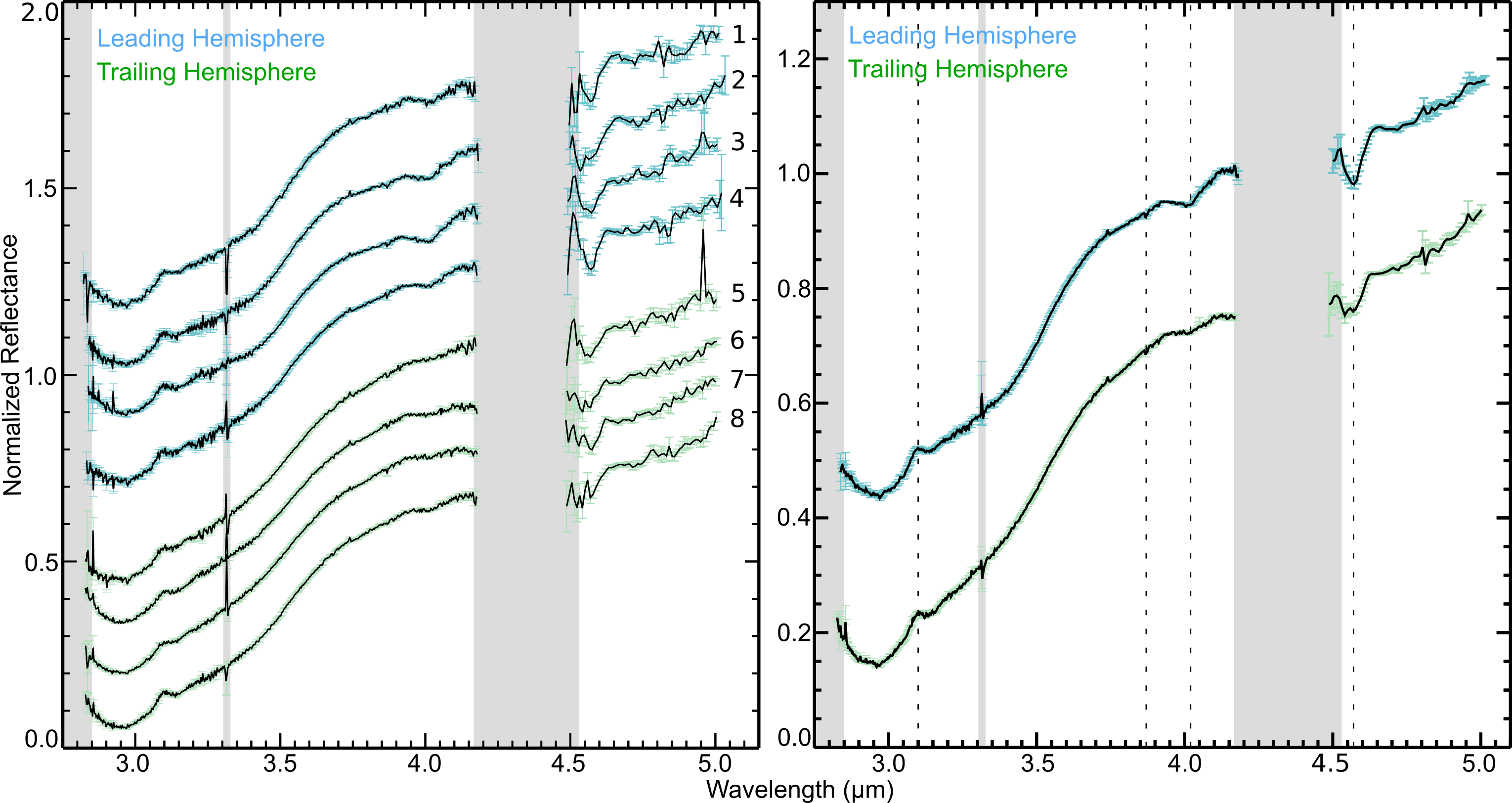}
	\caption{\textit{Left: Eight SpeX spectra of Callisto's leading and trailing hemisphere (1$\sigma$ uncertainties shown as blue and green error bars, respectively), normalized to 1 at 4.12 $\micron$ and offset vertically for clarity. The spectra are numbered (1 - 8) using the same sequence shown in Tables 1 and 2. Right: `Grand average' leading and trailing hemisphere spectra of Callisto (1$\sigma$ uncertainties shown as blue and green error bars, respectively), normalized to 1 at 4.12 $\micron$ and offset vertically for clarity. The dashed lines at 3.1, 3.87, 4.02, and 4.57 $\micron$ highlight the central wavelengths of the H$_2$O ice Fresnel peak and the 3.9-$\micron$, 4-$\micron$, and 4.6-$\micron$ bands, respectively. The 4-$\micron$ and 4.6-$micron$ bands are stronger on Callisto's leading hemisphere, whereas the 3.9-$\micron$ band does not display apparent hemispherical asymmetries in its band strength. Other, more subtle bands that are possibly present include features centered near 2.97, 3.05, and 3.4 $\micron$, which appear to be stronger on Callisto's leading hemisphere, and a band centered near 3.75 $\micron$, which appears to be slightly stronger on Callisto's trailing side (see Figure A4 for closer looks at these subtle bands). All spectra have been lightly smoothed using a 9 pixel-wide boxcar function between 2.75 and 4.2 $\micron$ and a 27 pixel-wide boxcar function between 4.45 and 5.05 $\micron$. Wavelength regions where strong telluric bands are present are shown as gray-toned zones. } }\vspace{0.1 cm}
\end{figure} 

 \begin{table}[htbp]
	\caption {4-$\micron$ band parameter measurements.} 
	\hskip1cm \begin{tabular}{*6c}
		\toprule	
			\begin{tabular}[c]{@{}l@{}}\hspace{-1 cm}Spectrum \\ \hspace{-1 cm}Number \end{tabular} & \begin{tabular}[c]{@{}l@{}}\hspace{-1 cm}Sub-Observer \\ \hspace{-1 cm}Longitude ($\degree$)\end{tabular} & \begin{tabular}[c]{@{}l@{}} \hspace{-1 cm}Sub-Observer \\  \hspace{-1 cm}Latitude  ($\degree$)\end{tabular} & \begin{tabular}[c]{@{}l@{}} \hspace{-1 cm}Longitudinal Quadrant \\ \hspace{-1 cm} (Geologic Region) \end{tabular} & 	\begin{tabular}[c]{@{}l@{}} \hspace{-1 cm}Band Area \\ \hspace{-1 cm}(10$^-$$^4$ $\micron$)   \end{tabular} & \begin{tabular}[c]{@{}l@{}} \hspace{-1 cm}Band Depth  \end{tabular} \\
		\midrule
		1 & 20.7& -1.0& Sub-Jupiter (Valhalla) &26.0 $\pm$  0.7& 0.027 $\pm$  0.002 \\
		2 & 64.0& -1.0&  Leading (Valhalla) &33.3 $\pm$  0.5& 0.033 $\pm$  0.001 \\
		3 & 120.6& -1.1& Leading (Asgard) &39.5 $\pm$  0.5& 0.039 $\pm$  0.001 \\
		4 & 164.7& -1.1& Anti-Jupiter (Asgard) &18.9 $\pm$  0.5& 0.021 $\pm$  0.001 \\
		5 & 200.7& -1.1& Anti-Jupiter &5.5 $\pm$  0.5& 0.007 $\pm$  0.001 \\
		6 & 244.0& -1.1&  Trailing &6.0 $\pm$  0.4& 0.008 $\pm$  0.001 \\
		7 & 287.2& -1.1&  Trailing &1.7 $\pm$  0.4& 0.007 $\pm$  0.001 \\
		8 & 337.8& -1.0& Sub-Jupiter &8.4 $\pm$  0.5& 0.012 $\pm$  0.001 \\
		\bottomrule
	\end{tabular}
\end{table}	

\subsection{Longitudinal distribution of the 4-$\micron$ band} 
Previous work has demonstrated that Callisto displays longitudinal trends in its surface composition, with weaker H$_2$O ice bands and more red material on its leading hemisphere \citep[e.g.,][]{morrison1974four}, and more CO$_2$ on its trailing hemisphere \citep[e.g.,][]{hibbitts2000distributions}. Consequently, we searched for longitudinal trends in the distribution of the 4-$\micron$ band on Callisto by calculating the mean 4-$\micron$ band areas for Callisto's leading and trailing hemispheres and comparing them: 29.43 $\pm$ 4.46 $\micron$$^{-4}$ and 5.40 $\pm$ 1.39 $\micron$$^{-4}$, respectively. We also calculated and compared the mean 4-$\micron$ band depths for Callisto's leading and trailing hemispheres: 0.030 $\pm$ 0.004 $\micron$ and 0.007 $\pm$ 0.001 $\micron$, respectively. Comparison of these values demonstrates that the mean 4-$\micron$ band is significantly stronger on Callisto's leading hemisphere compared to its trailing hemisphere ($>$4$\sigma$ difference).

To further investigate the distribution of the 4-$\micron$ band, we plotted the eight individual band area and depth measurements as a function of longitude (Figure 2). We fit the band area and depth measurements with a sinusoidal model, representing a surface displaying longitudinal trends in the distribution of the 4-$\micron$ band, and a mean model, representing a surface without leading/trailing trends in the distribution of this feature. We used an \textit{F}-test \citep[e.g.,][]{lomax2013introduction} to compare these two models. Our null hypothesis was that there is no difference between these two model fits, indicating that there is no meaningful longitudinal difference in the distribution of the 4-$\micron$ band. Our \textit{F}-test analysis demonstrates that there is a $>$3$\sigma$ difference between the mean and sinusoidal models for both the 4-$\micron$ band area and depth measurements (\textit{p} $<$0.002, Table A1). 

Thus, the mean 4-$\micron$ band areas and depths, as well as the results of our \textit{F}-test analysis, demonstrates that the 4-$\micron$ band is significantly stronger on Callisto's leading hemisphere. The 4-$\micron$ band is strongest in Spectrum 3 (mid-observation, sub-observer longitude 120.6$\degree$), which is located between the center of the Callisto's leading hemisphere (longitude 90$\degree$) and the Asgard impact basin (centered near 140$\degree$ longitude). Conversely, the 4-$\micron$ band is weakest in Spectrum 7 (mid-observation, sub-observer longitude of 287.2$\degree$), which is proximal to the center of Callisto's trailing side (longitude 270$\degree$). We consider the implications of the longitudinal distribution of the 4-$\micron$ band in Section 4.

\begin{figure}[h!]
 \centering
	\includegraphics[scale=0.70]{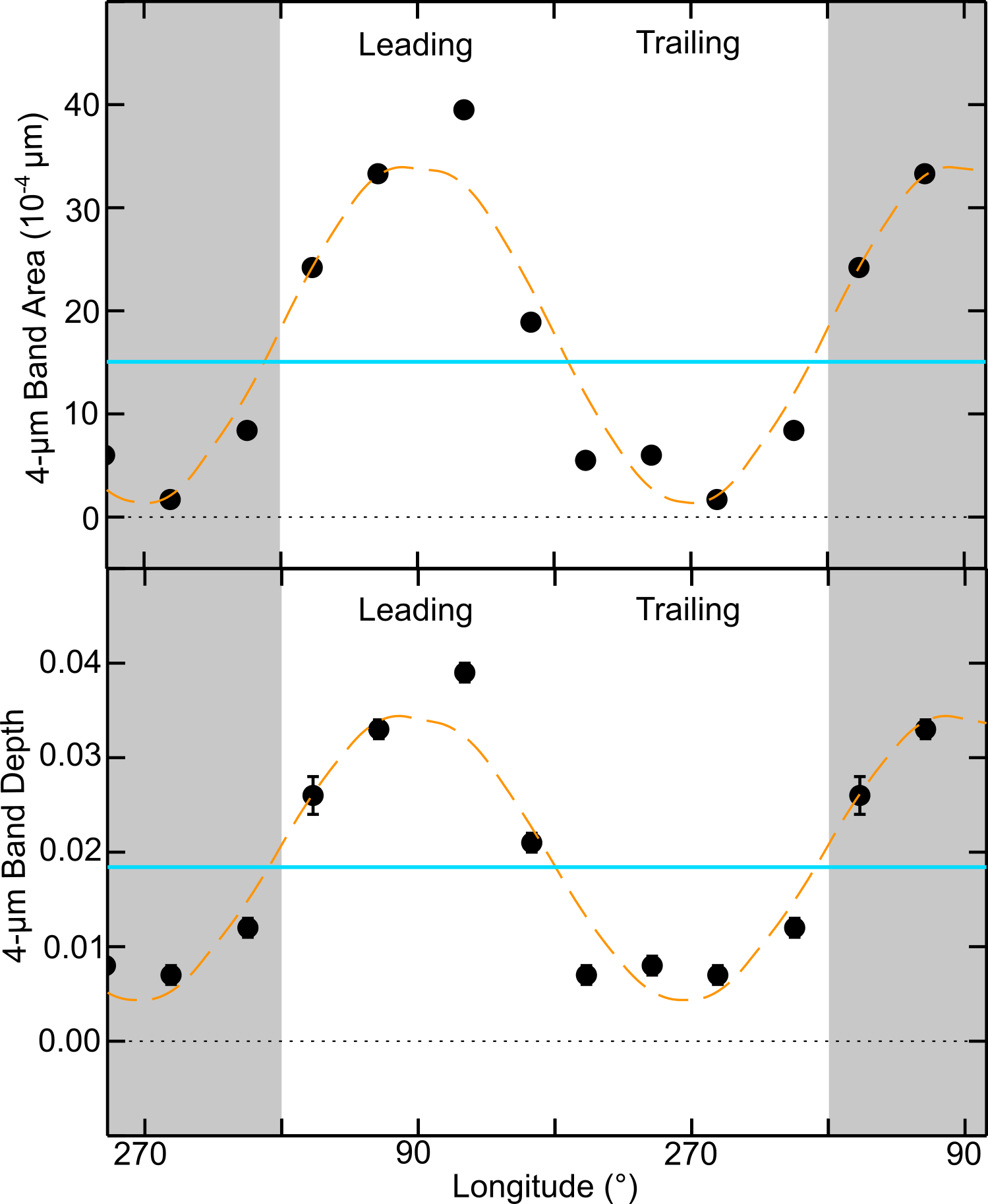}
	\caption{\textit{4-$\micron$ band area (top plot) and band depth (bottom plot) measurements and 1$\sigma$ uncertainties for all eight Callisto spectra, shown as a function of sub-observer longitude (Table 2). For the band area measurements, and some of the band depth measurements, the 1$\sigma$ uncertainties are smaller than the size of the black-filled circles. We fit these data with a sinusoidal model (orange dashed line), which represents a surface with significant variations in the abundance of this constituent, and the mean 4-$\micron$ band measurement (blue line), representing a surface that displays no variation in the abundance of this constituent. The maxima of the sinusoidal models are not locked to specific longitudes. Duplicate longitudes are shown to highlight the periodicity in the distribution of this constituent (gray-toned zones). We compared the model fits using an \textit{F}-test, finding that the 4-$\micron$ band displays significant longitudinal variations in its distribution (\textit{p} $<$0.002, F-test results summarized in Appendix A5). }}\vspace{0.1 cm}
\end{figure} 

\subsection{Comparing Callisto spectra to laboratory spectra of candidate species} 

To investigate the constituents contributing to the 4-$\micron$ band, we compared Callisto's grand average leading and trailing hemisphere spectra to laboratory reflectance spectra of candidate species (Figure 3). We first focused on sulfur-bearing materials that have been previously suggested (summarized in \citealt{moore2004callisto}), including: SO$_2$ mixed with H$_2$O ice (1:1 ratio, \citealt{moore2002ir}), SO$_2$ adsorbed on cabosil \citep{nash1995laboratory}, and thermally-altered sulfur \citep{fanale1979significance}. We also compared Callisto's 4-$\micron$ band to a spectral feature identified near 4.025 $\micron$, formed by irradiation of substrates composed of H$_2$O ice and hydrogen sulfide (H$_2$S) \citep{salama1990sulfur, moore2007radiolysis, jimenez2011sulfur}, which has been attributed to disulfanide (HS$_2$, \citealt{jimenez2011sulfur}). The 4.025 $\micron$ feature was extracted from a transmission (\textit{T}) spectrum calculated from absorbance (\textit{a}): \textit{T} = 10$^-$$^a$, measured at a temperature of 133 K \citep{jimenez2011sulfur}. Additionally, we compared Callisto's 4-$\micron$ band to laboratory reflectance spectra of the carbonates Na$_2$CO$_3$ and CaCO$_3$ \citep{nyquist1997handbook}, which have prominent 4-$\micron$ bands \citep{johnson2004radiation}. Previous studies have also considered carbonic acid, methanol, and other species, finding that these constituents do not provide good matches to the 4-$\micron$ band \citep[e.g.,][]{hudson2001radiation,hand2007energy}.

\begin{figure}
	\centering
	\includegraphics[scale=0.70]{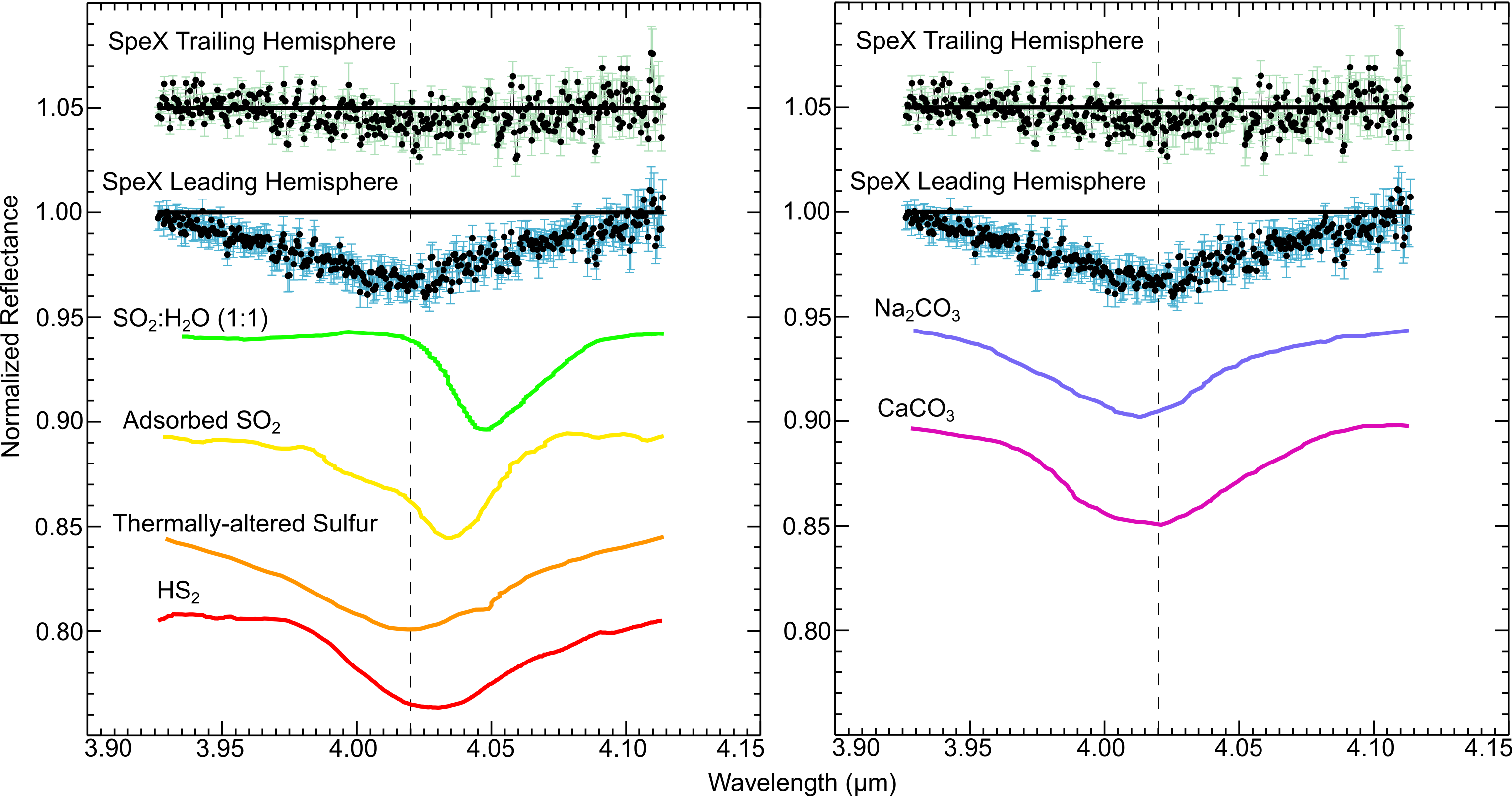}
	\caption{\textit{Left: Comparison between continuum-divided 4-$\micron$ bands identified in the grand average leading and trailing hemisphere spectra of Callisto (shown in Figure 1) and continuum-divided laboratory spectra of SO$_2$ in H$_2$O ice (green, \citealt{moore2002ir}), adsorbed SO$_2$ (gold, \citealt{nash1995laboratory}), thermally-altered S (orange, \citealt{fanale1979significance}), and a transmission feature centered near 4.025 $\micron$ (red) that has been attributed to HS$_2$ \citep{jimenez2011sulfur}, offset vertically for clarity. Right: Comparison between Callisto's continuum-divided 4-$\micron$ bands and continuum-divided laboratory spectra of Na$_2$CO$_3$ and CaCO$_3$ (violet and purple, respectively, \citealt{nyquist1997handbook}), offset vertically for clarity. In both plots, the dash line at 4.02 $\micron$ marks the central wavelength position of the 4-$\micron$ band identified in the SpeX spectra of Callisto.}}\vspace{0.1 cm}
\end{figure} 

Our comparison demonstrates that SO$_2$-bearing materials are poor matches to Callisto's 4-$\micron$ band, both in terms of band center and shape (Figure 3).  In contrast, thermally-altered sulfur provides a reasonable match in terms of band center and shape. The spectrum shown in Figure 3 is from a sample heated to $\sim$473 K to produce polymeric sulfur, which likely includes S$_2$, S$_3$, or other S allotropes (Fanale et al., 1979). Although this laboratory sample was formed at substantially higher temperatures than Callisto’s estimated peak surface temperature ($\sim$165 K, \citealt{spencer1987surfaces}), S allotropes could also be formed by radiolytic modification of Callisto surface. For example, Iogenic S ions delivered to Callisto in Jupiter’s magnetosphere should react with its surface constituents, spurring a radiolytic production cycle that could generate S allotropes, similar to the sulfur cycle operating on Europa  \citep[e.g.,][]{carlson2002sulfuric}. 

The 4.025 $\micron$ feature, attributed to HS$_2$, also provides a reasonable match to the band center and shape of Callisto's 4-$\micron$ band. This 4.025 $\micron$ feature was formed in the laboratory by irradiating mixtures of H$_2$O ice and H$_2$S at cryogenic temperatures ($<$ 90 K). After irradiation, these samples were warmed up to 224 K, and the 4.025 $\micron$ feature persisted over the entire temperature range, whereas the H$_2$S frost fundamental band centered near 3.9 $\micron$ ($\nu$$_1$) started to decrease once the sample chamber reached temperatures $>$ 100 K (see Figure 6 in \citealt{jimenez2011sulfur}). Therefore, at Callisto's peak surface temperature ($\sim$165 K),  H$_2$S should sublimate, thereby weakening the 3.9-$\micron$ H$_2$S fundamental band, whereas the 4.025 $\micron$ feature attributed to HS$_2$ should persist, possibly explaining the weakness of the 3.9-$\micron$ band and stronger 4-$\micron$ band we detected with SpeX. Of note, UV irradiation of H$_2$S ice should also lead to the formation of S allotropes like S$_2$ and S$_3$ \citep{jimenez2011sulfur}. 

Carbonates like Na$_2$CO$_3$ and CaCO$_3$ also provide good matches to the 4-$\micron$ feature in terms of central wavelength position and band shape, as previously demonstrated \citep{johnson2004radiation}. However, carbonates display prominent absorption features at shorter wavelengths (between $\sim$2.2 - 2.4 and $\sim$3.3 - 3.5 $\micron$), which we do not detect in the SpeX spectra of Callisto (Figures 1 and A2) and were not detected by NIMS, reducing the likelihood that carbonates are the primary contributors to the 4-$\micron$ band. Nevertheless, the surface of Callisto has a vigorous carbon system, and perhaps C- and S-bearing species, along with H$_2$O ice, participate in a shared radiolytic cycle, similar to sulfur and H$_2$O ice on Europa \citep[e.g.,][]{carlson2002sulfuric}. Such a C-S-H$_2$O radiolytic cycle might produce species like carbonyl sulfide (OCS) and carbon disulfide (CS$_2$), which display absorption features at 3.4 and 4.6 $\micron$, respectively \citep{CoblentzSoc1964}, possibly explaining the presence of these two features in NIMS and SpeX spectra of Callisto. 

In summary, we find that thermally-altered sulfur provides a good match to the spectral signature of Callisto's 4-$\micron$ band. The 4.025 $\micron$ feature attributed to HS$_2$ can also explain the presence of the 4-$\micron$ band. Thus, our analysis supports the presence of S-bearing constituents on Callisto, but we do not find strong evidence for the previously suggested SO$_2$. 

\section{Discussion and Conclusions} 
We analyzed eight SpeX spectra of Callisto, gaining new knowledge on the spectral signature and distribution of the 4-$\micron$ band. We identified a leading/trailing hemispherical trend in the distribution of the 4-$\micron$ band, with significantly stronger band measurements on Callisto's leading side. Our analysis suggests that the spectral signature of Callisto's 4-$\micron$ band is consistent with thermally-altered sulfur and a feature centered near 4.025 $\micron$, which has been attributed to HS$_2$ \citep{jimenez2011sulfur}. Below, we consider possible origin scenarios to explain the presence of the 4-$\micron$ band on Callisto.

\textit{Delivery of S-bearing dust from the irregular satellites:} Dust grains ($\sim$10 - 1000 $\micron$ diameters) ejected from the giant planets' retrograde irregular satellites should experience Poynting-Robertson drag and slowly migrate inward on decaying orbits  \citep[e.g.,][]{burns1979radiation}. Eventually, the orbits of these dust grains overlap the orbital zone of the classical satellites, and they subsequently collide with the leading hemispheres of the outermost moons \citep[e.g.,][]{bottke2013black, tamayo2013chaotic}. The irregular satellites of the giant planets are darker and spectrally redder than the classical moons \citep[e.g.,][]{graykowski2018colors}. The accumulation of irregular satellite dust can therefore explain why the leading hemispheres of the outer classical moons Callisto \citep[e.g.,][]{pollack1978near}, Iapetus \citep[e.g.][]{cruikshank1983dark}, and Titania and Oberon \citep[e.g.,][]{cartwright2018red} are spectrally redder and darker than their trailing hemispheres. 

Similar to the irregular satellites, Jupiter's trojan asteroid population includes a group of objects with reddish surfaces \citep[e.g.,][]{szabo2007properties,emery2010near}. It has been hypothesized that this red-colored trojan asteroid group is populated by captured Kuiper Belt Objects (KBOs) that formed beyond an H$_2$S `snow line' in the primordial Kuiper Belt \citep{wong2016hypothesis}. Dynamical simulations indicate that giant planet migration in the early Solar System scattered large numbers of KBOs into the giant planet zone, and some of these objects were likely captured by Jupiter into its trojan asteroid and irregular satellite populations \citep[e.g.,][]{morbidelli2020kuiper}. Along with capture of H$_2$S-rich KBOs in Jupiter's L4 and L5 Lagrange points, perhaps H$_2$S-bearing KBOs were captured into Jupiter’s irregular satellite population, providing a source of H$_2$S and other S-bearing species that could be delivered to Callisto's leading hemisphere within dust grains. Although sublimation and space weathering should remove volatile-rich deposits from the surfaces of Jovian irregular satellites, H$_2$S and other volatiles could be retained beneath their regoliths and subsequently excavated and ejected in dust grains by impact events. Albeit, it is uncertain how long H$_2$S could survive within dust grains before colliding with Callisto, nor whether H$_2$S delivered to Callisto would survive long enough to be radiolytically modified into other S-bearing species. Laboratory work that explores the longevity of H$_2$S under conditions relevant to the surface of Callisto are needed to further investigate this scenario. 

\textit{Delivery of magnetospheric S ions:} Volcanoes on Io erupt large volumes of S-rich material into orbit. Much of the erupted sulfur is then ionized and gets trapped in Jupiter's magnetosphere \citep[e.g.,][]{schneider2007io}. These trapped S ions are delivered to the more distant icy Galilean moons, primarily bombarding the trailing hemispheres of Europa and Ganymede \citep[e.g.,][]{johnson2004radiation}. In contrast, recent numerical modeling work shows that energetic O and S ions (1 - 100 KeV) trapped in Jupiter's magnetosphere might preferentially bombard Callisto's leading hemisphere \citep{liuzzo2019energetic}. This hemispherical dichotomy results from interactions between Jupiter's magnetosphere and Callisto's intense ionosphere \citep{kliore2002ionosphere}, which largely blocks these lower energy O and S ions from accessing its trailing side, unlike Europa and Ganymede, which do not have substantial ionospheres. Radiolytic sulfur chemistry, induced by implanted S ions, could therefore be operating across Callisto's leading hemisphere, explaining the stronger 4-$\micron$ band we detected on this hemisphere. 

\textit{Exposure of native S-rich deposits:} Alternatively, the species contributing to the 4-$\micron$ band could be native to Callisto and are exposed by dust-driven regolith overturn, which should preferentially operate on the leading hemispheres of tidally-locked moons \citep[e.g.,][]{bennett2013space}. In this scenario, native deposits of S-bearing species exposed by dust collisions could  be modified by UV photolysis and charged particle radiolysis, forming new S-rich constituents, such as S$_2$-bearing species and perhaps CS-bearing species as well. Larger impact events might also excavate S-rich deposits from greater depths in Callisto's subsurface, albeit, analysis of NIMS data indicates that the 4-$\micron$ band is not spatially associated with craters on Callisto's leading hemisphere, unlike CO$_2$ \citep{hibbitts2000distributions}. 

To further test these different origin hypotheses, follow up ground-based observations of the 4-$\micron$ band at complementary sub-observer longitudes are needed. Furthermore, new laboratory experiments that measure the spectral signature of substrates composed of H$_2$O ice, S-bearing, and C-bearing species, performed under temperature and irradiation conditions relevant to the surface of Callisto, could provide key insight into origin and composition of the 4-$\micron$ band. These new experiments could also provide insight into whether a radiolytic cycle involving S, C, and H$_2$O ice is occurring on Callisto, spurring the formation of species like OCS and CS$_2$.
 
\section{Acknowledgments} 
The observations reported here were made from the summit of Maunakea, and we thank the people of Hawaii for the opportunity to observe from this special mountain. Kevin Hand and Tom Nordheim acknowledge the support of the Jet Propulsion Laboratory, California Institute of Technology, under contract with NASA. Kevin Hand also acknowledges the support of NASA's `Exploring Ocean Worlds' Research Coordinated Network, managed by the Woods Hole Oceanographic Institution. 

\bibliography{references}{}

\begin{thebibliography}{}
\expandafter\ifx\csname natexlab\endcsname\relax\def\natexlab#1{#1}\fi
\providecommand{\url}[1]{\href{#1}{#1}}
\providecommand{\dodoi}[1]{doi:~\href{http://doi.org/#1}{\nolinkurl{#1}}}
\providecommand{\doeprint}[1]{\href{http://ascl.net/#1}{\nolinkurl{http://ascl.net/#1}}}
\providecommand{\doarXiv}[1]{\href{https://arxiv.org/abs/#1}{\nolinkurl{https://arxiv.org/abs/#1}}}

\bibitem[{Bennett {et~al.}(2013)Bennett, Pirim, \& Orlando}]{bennett2013space}
Bennett, C.~J., Pirim, C., \& Orlando, T.~M. 2013, Chemical reviews, 113, 9086

\bibitem[{Bottke {et~al.}(2013)Bottke, Vokrouhlick{\`y}, Nesvorn{\`y}, \&
  Moore}]{bottke2013black}
Bottke, W.~F., Vokrouhlick{\`y}, D., Nesvorn{\`y}, D., \& Moore, J.~M. 2013,
  Icarus, 223, 775

\bibitem[{Burns {et~al.}(1979)Burns, Lamy, \& Soter}]{burns1979radiation}
Burns, J.~A., Lamy, P.~L., \& Soter, S. 1979, Icarus, 40, 1

\bibitem[{Calvin \& Clark(1993)}]{calvin1993spectral}
Calvin, W.~M., \& Clark, R.~N. 1993, Icarus, 104, 69

\bibitem[{Carlson {et~al.}(2002)Carlson, Anderson, Johnson, Schulman, \&
  Yavrouian}]{carlson2002sulfuric}
Carlson, R., Anderson, M., Johnson, R., Schulman, M., \& Yavrouian, A. 2002,
  Icarus, 157, 456

\bibitem[{Carlson {et~al.}(1996)Carlson, Smythe, Baines, Barbinis, Becker,
  Burns, Calcutt, Calvin, Clark, Danielson, {et~al.}}]{carlson1996near}
Carlson, R., Smythe, W., Baines, K., {et~al.} 1996, Science, 274, 385

\bibitem[{Cartwright {et~al.}(2020{\natexlab{a}})Cartwright, Emery, Grundy,
  Cruikshank, Beddingfield, \& Pinilla-Alonso}]{cartwright2020probing}
Cartwright, R.~J., Emery, J.~P., Grundy, W.~M., {et~al.} 2020{\natexlab{a}},
  Icarus, 338, 113513

\bibitem[{Cartwright {et~al.}(2018)Cartwright, Emery, Pinilla-Alonso, Lucas,
  Rivkin, \& Trilling}]{cartwright2018red}
Cartwright, R.~J., Emery, J.~P., Pinilla-Alonso, N., {et~al.} 2018, Icarus,
  314, 210

\bibitem[{Cartwright {et~al.}(2015)Cartwright, Emery, Rivkin, Trilling, \&
  Pinilla-Alonso}]{cartwright2015distribution}
Cartwright, R.~J., Emery, J.~P., Rivkin, A.~S., Trilling, D.~E., \&
  Pinilla-Alonso, N. 2015, Icarus, 257, 428

\bibitem[{Cartwright {et~al.}(2020{\natexlab{b}})Cartwright, Beddingfield,
  Nordheim, Roser, Grundy, Hand, Emery, Cruikshank, \&
  Scipioni}]{cartwright2020evidence}
Cartwright, R.~J., Beddingfield, C.~B., Nordheim, T.~A., {et~al.}
  2020{\natexlab{b}}, The Astrophysical Journal Letters, 898, L22

\bibitem[{Coblentz-Society-Inc.(1964)}]{CoblentzSoc1964}
Coblentz-Society-Inc. 1964, NIST Chemistry WebBook, NIST Standard Reference
  Database, SRD 69

\bibitem[{Cruikshank {et~al.}(1983)Cruikshank, Bell, Gaffey, Brown, Howell,
  Beerman, \& Rognstad}]{cruikshank1983dark}
Cruikshank, D., Bell, J., Gaffey, M., {et~al.} 1983, Icarus, 53, 90

\bibitem[{Cruikshank(1980)}]{cruikshank1980infrared}
Cruikshank, D.~P. 1980, Icarus, 41, 240

\bibitem[{Cushing {et~al.}(2004)Cushing, Vacca, \&
  Rayner}]{cushing2004spextool}
Cushing, M.~C., Vacca, W.~D., \& Rayner, J.~T. 2004, Publications of the
  Astronomical Society of the Pacific, 116, 362

\bibitem[{Emery {et~al.}(2010)Emery, Burr, \& Cruikshank}]{emery2010near}
Emery, J.~P., Burr, D.~M., \& Cruikshank, D.~P. 2010, The Astronomical Journal,
  141, 25

\bibitem[{Fanale {et~al.}(1979)Fanale, Brown, Cruikshank, \&
  Clake}]{fanale1979significance}
Fanale, F.~P., Brown, R.~H., Cruikshank, D.~P., \& Clake, R.~N. 1979, Nature,
  280, 761

\bibitem[{Graykowski \& Jewitt(2018)}]{graykowski2018colors}
Graykowski, A., \& Jewitt, D. 2018, The Astronomical Journal, 155, 184

\bibitem[{Hage {et~al.}(1998)Hage, Liedl, Hallbrucker, \&
  Mayer}]{hage1998carbonic}
Hage, W., Liedl, K.~R., Hallbrucker, A., \& Mayer, E. 1998, Science, 279, 1332

\bibitem[{Hand {et~al.}(2007)Hand, Carlson, \& Chyba}]{hand2007energy}
Hand, K.~P., Carlson, R.~W., \& Chyba, C.~F. 2007, Astrobiology, 7, 1006

\bibitem[{Hendrix \& Johnson(2008)}]{hendrix2008callisto}
Hendrix, A.~R., \& Johnson, R.~E. 2008, The Astrophysical Journal, 687, 706

\bibitem[{Hendrix {et~al.}(2019)Hendrix, Hurford, Barge, Bland, Bowman,
  Brinckerhoff, Buratti, Cable, Castillo-Rogez, Collins,
  {et~al.}}]{hendrix2019nasa}
Hendrix, A.~R., Hurford, T.~A., Barge, L.~M., {et~al.} 2019, Astrobiology, 19,
  1

\bibitem[{Hibbitts {et~al.}(2002)Hibbitts, Klemaszewski, McCord, Hansen, \&
  Greeley}]{hibbitts2002co2}
Hibbitts, C., Klemaszewski, J., McCord, T., Hansen, G., \& Greeley, R. 2002,
  Journal of Geophysical Research: Planets, 107, 14

\bibitem[{Hibbitts {et~al.}(2000)Hibbitts, McCord, \&
  Hansen}]{hibbitts2000distributions}
Hibbitts, C., McCord, T., \& Hansen, G. 2000, Journal of Geophysical Research:
  Planets, 105, 22541

\bibitem[{Hudson \& Moore(2001)}]{hudson2001radiation}
Hudson, R., \& Moore, M. 2001, Journal of Geophysical Research: Planets, 106,
  33275

\bibitem[{Jim{\'e}nez-Escobar \& Mu{\~n}oz~Caro(2011)}]{jimenez2011sulfur}
Jim{\'e}nez-Escobar, A., \& Mu{\~n}oz~Caro, G. 2011, Astronomy \& Astrophysics,
  536, A91

\bibitem[{Johnson {et~al.}(2004)Johnson, Carlson, Cooper, Paranicas, Moore, \&
  Wong}]{johnson2004radiation}
Johnson, R., Carlson, R., Cooper, J., {et~al.} 2004, Jupiter: The Planet,
  Satellites and Magnetosphere, 485

\bibitem[{Kliore {et~al.}(2002)Kliore, Anabtawi, Herrera, Asmar, Nagy, Hinson,
  \& Flasar}]{kliore2002ionosphere}
Kliore, A., Anabtawi, A., Herrera, R., {et~al.} 2002, Journal of Geophysical
  Research: Space Physics, 107, SIA

\bibitem[{Lane \& Domingue(1997)}]{lane1997iue}
Lane, A.~L., \& Domingue, D.~L. 1997, Geophysical research letters, 24, 1143

\bibitem[{Liuzzo {et~al.}(2019)Liuzzo, Simon, \& Regoli}]{liuzzo2019energetic}
Liuzzo, L., Simon, S., \& Regoli, L. 2019, Planetary and Space Science, 166, 23

\bibitem[{Lomax \& Hahs-Vaughn(2013)}]{lomax2013introduction}
Lomax, R.~G., \& Hahs-Vaughn, D.~L. 2013, An introduction to statistical
  concepts (Routledge)

\bibitem[{Mastrapa {et~al.}(2009)Mastrapa, Sandford, Roush, Cruikshank, \&
  Dalle~Ore}]{mastrapa2009optical}
Mastrapa, R., Sandford, S., Roush, T., Cruikshank, D., \& Dalle~Ore, C. 2009,
  The Astrophysical Journal, 701, 1347

\bibitem[{McCord {et~al.}(1997)McCord, Carlson, Smythe, Hansen, Clark,
  Hibbitts, Fanale, Granahan, Segura, Matson, {et~al.}}]{mccord1997organics}
McCord, T.~a., Carlson, R., Smythe, W., {et~al.} 1997, Science, 278, 271

\bibitem[{McCord {et~al.}(1998)McCord, Hansen, Clark, Martin, Hibbitts, Fanale,
  Granahan, Segura, Matson, Johnson, {et~al.}}]{mccord1998non}
McCord, T.~a., Hansen, G., Clark, R., {et~al.} 1998, Journal of Geophysical
  Research: Planets, 103, 8603

\bibitem[{Moore {et~al.}(2004)Moore, Chapman, Bierhaus, Greeley, Chuang,
  Klemaszewski, Clark, Dalton, Hibbitts, Schenk, {et~al.}}]{moore2004callisto}
Moore, J.~M., Chapman, C.~R., Bierhaus, E.~B., {et~al.} 2004, jpsm, 1, 397

\bibitem[{Moore {et~al.}(2002)Moore, Hudson, \& Carlson}]{moore2002ir}
Moore, M., Hudson, R., \& Carlson, R. 2002, in DPS, Vol.~34, 35--02

\bibitem[{Moore {et~al.}(2007)Moore, Hudson, \& Carlson}]{moore2007radiolysis}
Moore, M., Hudson, R., \& Carlson, R. 2007, Icarus, 189, 409

\bibitem[{Morbidelli \& Nesvorn{\`y}(2020)}]{morbidelli2020kuiper}
Morbidelli, A., \& Nesvorn{\`y}, D. 2020, in The Trans-Neptunian Solar System
  (Elsevier), 25--59

\bibitem[{Morrison {et~al.}(1974)Morrison, Morrison, \&
  Lazarewicz}]{morrison1974four}
Morrison, D., Morrison, N.~D., \& Lazarewicz, A.~R. 1974, Icarus, 23, 399

\bibitem[{Nash \& Betts(1995)}]{nash1995laboratory}
Nash, D.~B., \& Betts, B.~H. 1995, Icarus, 117, 402

\bibitem[{Noll {et~al.}(1997)Noll, Johnson, McGrath, \&
  Caldwell}]{noll1997detection}
Noll, K.~S., Johnson, R.~E., McGrath, M.~A., \& Caldwell, J.~J. 1997,
  Geophysical research letters, 24, 1139

\bibitem[{Nyquist {et~al.}(1997)Nyquist, Leugers, \&
  Putzig}]{nyquist1997handbook}
Nyquist, R.~A., Leugers, M.~A., \& Putzig, C.~L. 1997, The handbook of infrared
  and raman spectra of inorganic compounds and organic salts:(a 4-volume set),
  Acad. Press

\bibitem[{Pollack {et~al.}(1978)Pollack, Witteborn, Erickson, Strecker,
  Baldwin, \& Bunch}]{pollack1978near}
Pollack, J.~B., Witteborn, F.~C., Erickson, E.~F., {et~al.} 1978, Icarus, 36,
  271

\bibitem[{Rayner {et~al.}(2003)Rayner, Toomey, Onaka, Denault, Stahlberger,
  Vacca, Cushing, \& Wang}]{rayner2003spex}
Rayner, J., Toomey, D., Onaka, P., {et~al.} 2003, Publications of the
  Astronomical Society of the Pacific, 115, 362

\bibitem[{Salama {et~al.}(1990)Salama, Allamandola, Witteborn, Cruikshank,
  Sandford, \& Bregman}]{salama1990sulfur}
Salama, F., Allamandola, L., Witteborn, F., {et~al.} 1990, Icarus, 83, 66

\bibitem[{Schneider \& Bagenal(2007)}]{schneider2007io}
Schneider, N.~M., \& Bagenal, F. 2007, in Io After Galileo (Springer), 265--286

\bibitem[{Spencer(1987)}]{spencer1987surfaces}
Spencer, J.~R. 1987, Retrieved from:
  https://repository.arizona.edu/handle/10150/184098

\bibitem[{Szab{\'o} {et~al.}(2007)Szab{\'o}, Ivezi{\'c}, Juri{\'c}, \&
  Lupton}]{szabo2007properties}
Szab{\'o}, G.~M., Ivezi{\'c}, {\v{Z}}., Juri{\'c}, M., \& Lupton, R. 2007,
  Monthly Notices of the Royal Astronomical Society, 377, 1393

\bibitem[{Tamayo {et~al.}(2013)Tamayo, Burns, \& Hamilton}]{tamayo2013chaotic}
Tamayo, D., Burns, J.~A., \& Hamilton, D.~P. 2013, Icarus, 226, 655

\bibitem[{Trumbo {et~al.}(2017)Trumbo, Brown, Fischer, \& Hand}]{trumbo2017new}
Trumbo, S.~K., Brown, M.~E., Fischer, P.~D., \& Hand, K.~P. 2017, The
  Astronomical Journal, 153, 250

\bibitem[{Wong \& Brown(2016)}]{wong2016hypothesis}
Wong, I., \& Brown, M.~E. 2016, The Astronomical Journal, 152, 90

\end{thebibliography}
\bibliographystyle{aasjournal}



\renewcommand{\thesection}{\arabic{section}}

\renewcommand{\thesubsection}{A\arabic{subsection}}
\setcounter{figure}{0}

\renewcommand{\thefigure}{A\arabic{figure}}
\setcounter{figure}{0}

\renewcommand{\thetable}{A\arabic{table}}
\setcounter{table}{0}

\centering\textbf{Appendix}\vspace{-0.0cm}

\subsection{Methods: Band parameter measurements}

\justify
In this appendix, we show an example of our band parameter measurement technique (Figure A1).\vspace{-0.0cm}
	
\begin{figure}[h!]
	\centering
	\includegraphics[scale=0.70]{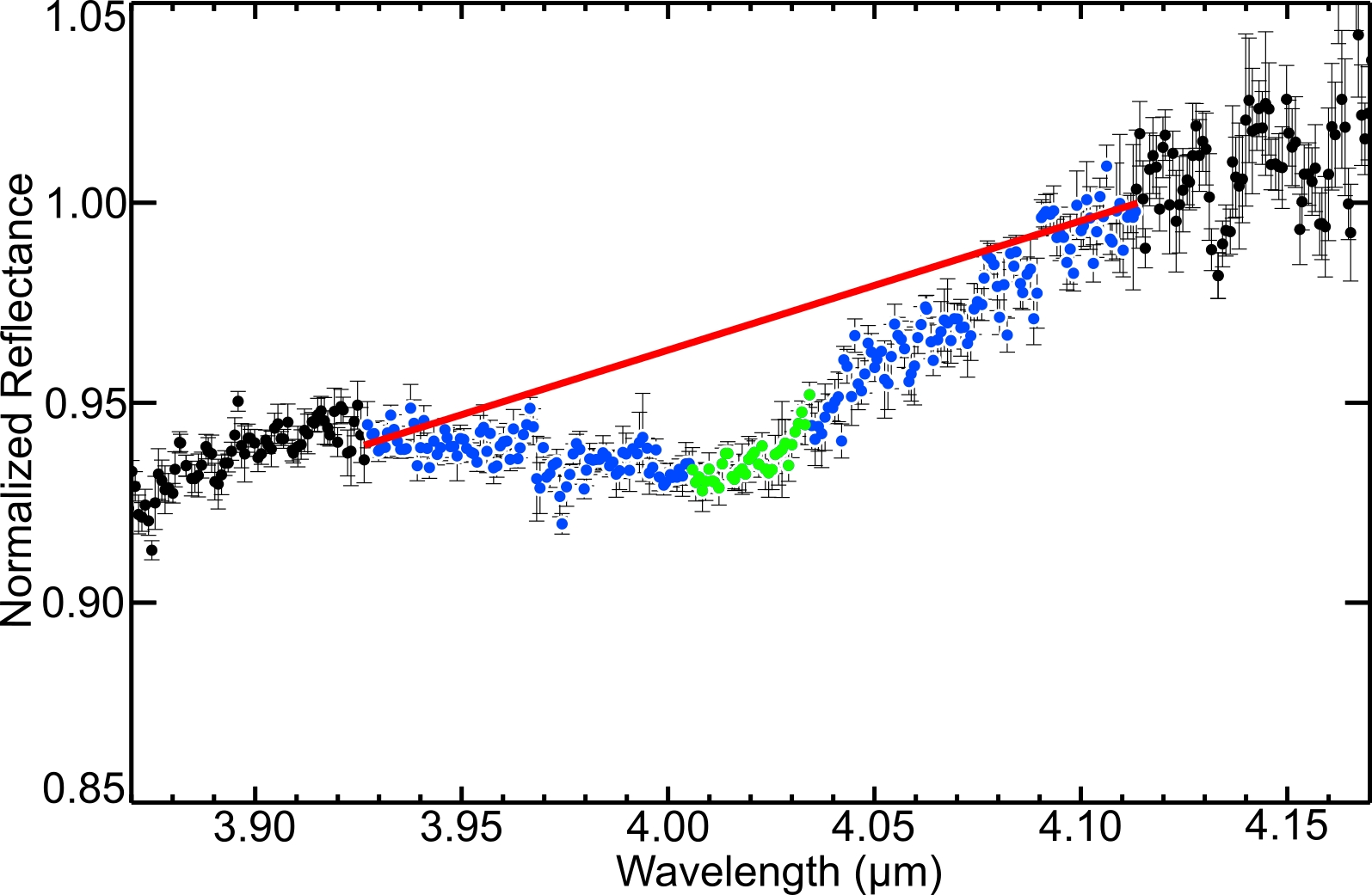}
	\caption{\textit{An example of our 4-$\micron$ band area and depth measurement procedure, using a SpeX spectrum of Callisto collected at mid-observation, sub-observer longitude 120.6$\degree$ (Spectrum 3 in Figure 1 and Tables 1 and 2). The data points used for our band area measurement (blue, 3.926 -- 4.114 $\micron$) and those used for our band depth measurement (green, 4.005 -- 4.035 $\micron$) are highlighted. The 4-$\micron$ band continuum was calculated by averaging the data points within $\pm$ 0.01 $\micron$ of 3.926 and 4.114 $\micron$, and we then connected these two mean continuum reflectances with a line (red).}}
\end{figure}

\newpage

\subsection{Results: IRTF/SpeX spectra}

\justify
In this appendix, we present eight near-infrared reflectance spectra of the Galilean moon Callisto at their native spectral resolution (Figure A2). \vspace{-0.0cm}
	
\begin{figure}[h!]
	\centering
	\includegraphics[scale=0.90]{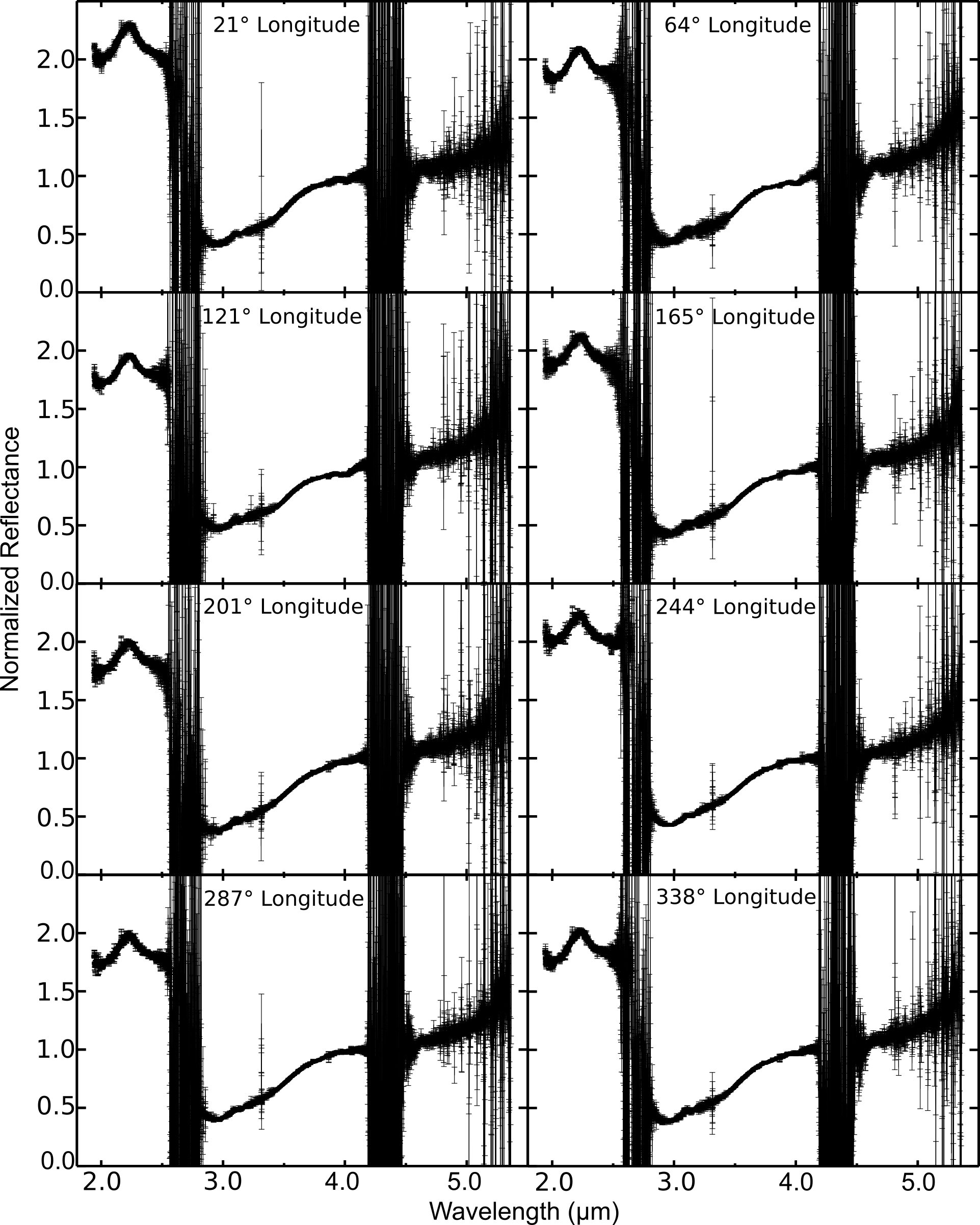}
	\caption{\textit{Eight new SpeX spectra of Callisto and their 1$\sigma$ uncertainties, collected in May and June, 2020. The mid-observation, sub-observer longitude for each spectrum is listed along the top of each plot (see Table 1 for observation details). All spectra have been normalized to 1 at 4.12 $\micron$.}}
\end{figure}

\subsection{Results: Comparison of 4-$\micron$ bands in SpeX and NIMS spectra}

\justify
In this appendix, we compare continuum-divided 4-$\micron$ bands detected in the grand average leading and trailing hemisphere SpeX spectra to average spectra extracted from three different NIMS observations of Callisto (Figure A3).

\begin{figure}[h!]
	\centering
	\includegraphics[scale=0.70]{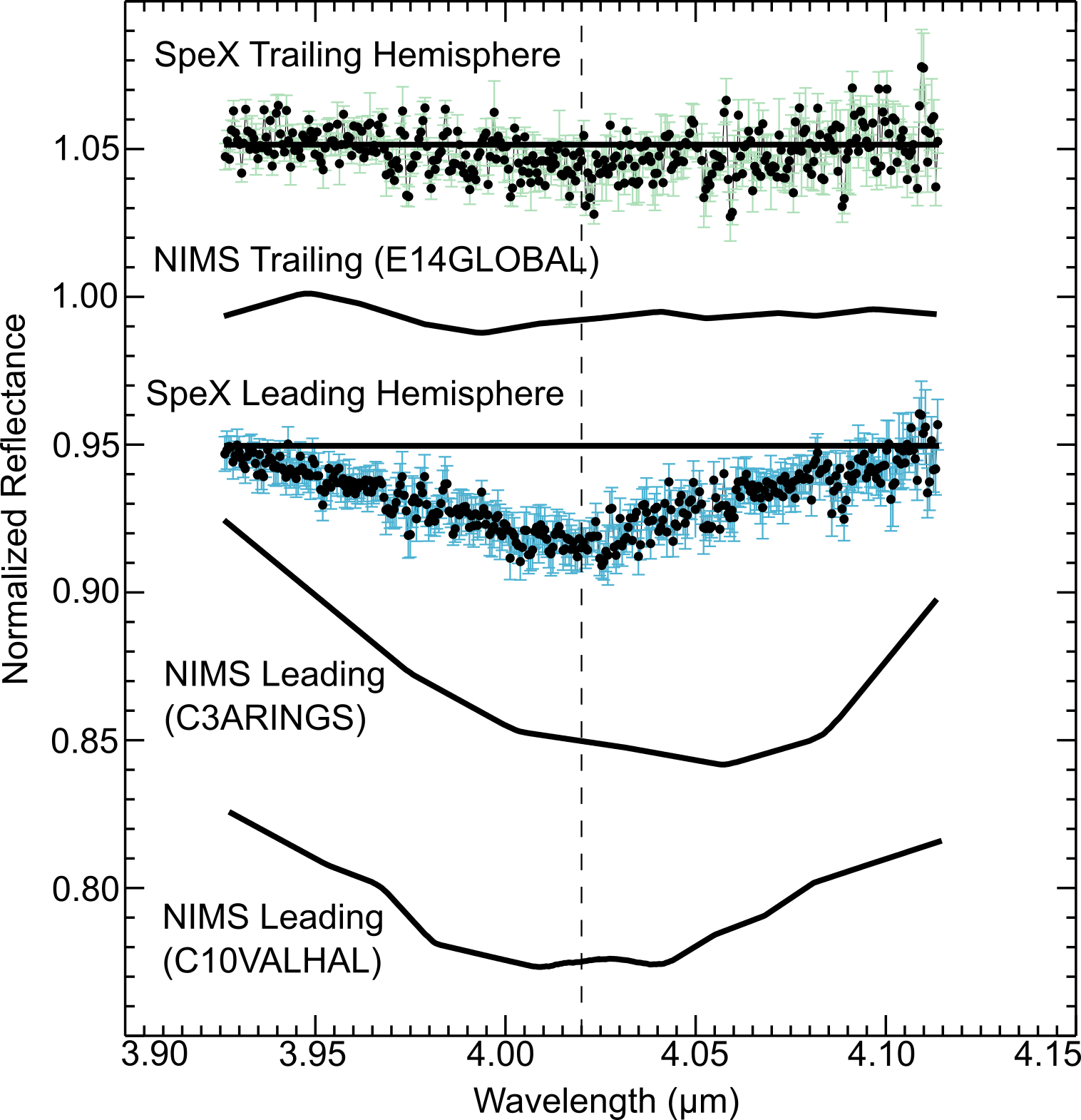}
	\caption{\textit{Comparison between continuum-divided 4-$\micron$ bands observed in the grand average SpeX spectra of Callisto's leading and trailing hemispheres (blue and green error bars, respectively) to continuum-divided 4-$\micron$ bands observed  in average NIMS spectra collected over Callisto's trailing hemisphere (E14GLOBAL), proximal to the Asgard impact basin (C3ARINGS), and proximal to the Valhalla impact basin (C10VALHAL), offset vertically for clarity. The dash line at 4.02 $\micron$ marks the central wavelength position of the 4-$\micron$ band identified in the SpeX spectra of Callisto. }}
\end{figure}

\subsection{Results: Subtle absorption bands in the grand average IRTF/SpeX spectra}

\justify
In this appendix, we present grand average Callisto spectra, zoomed in to the $\sim$2.85 - 4.15 $\micron$ wavelength region, to highlight some of the subtle absorption features hinted at in these data (Figure A4). The grand average spectra display subtle bands centered near 2.97 $\micron$, which appears to be stronger in the mean leading hemisphere spectrum. This band was previously detected by NIMS, and attributed to trapped OH \citep[e.g.,][]{moore2004callisto}. Alternatively, the N-H asymmetric stretch fundamental band ($\nu$$_3$) is centered near 2.97 $\micron$ and could explain the presence of this feature as well. The leading hemisphere spectrum also displays a subtle band at 3.05 $\micron$, which was originally detected by prior ground-based observations and attributed to the presence of ammonium-bearing clays \citep[e.g.,][]{calvin1993spectral}. We also note a weak band at 3.4 $\micron$ on Callisto's leading hemisphere, which was detected previously by ground-based observers and attributed to the presence of small H$_2$O ice grains \citep[e.g.,][]{calvin1993spectral}, as well as by NIMS, attributed to the presence of organics \citep[e.g.,][]{mccord1998non}. This subtle 3.4 $\micron$ feature could also result from carbonyl sulfide (OCS, \citealt{CoblentzSoc1964}). The detection of these possible N-bearing features at 2.97 and 3.05 $\micron$, and a possible C- and S-bearing band at 3.4 $\micron$, primarily on Callisto's leading hemisphere, is consistent with the stronger 4.6-$\micron$ band on Callisto's leading side (Figure 1). The  4.6-$\micron$ band has been attributed to CN-bearing species by some studies \citep[e.g.,][]{mccord1998non}, and we suggest that it could also stem from carbon disulfide (CS$_2$, \citealt{CoblentzSoc1964}). Similar to our preferred hypotheses for the origin of the 4-$\micron$ band, perhaps the features centered near 2.97, 3.05, 3.4, and 4.6 $\micron$ result from constituents delivered in dust grains from the retrograde irregular satellites. Alternatively, perhaps the  features near 3.4 and 4.6 $\micron$ result from radiolytic products that are formed by magnetospheric charged particles bombarding Callisto's surface.

 We also note the presence of a subtle band centered near 3.75 $\micron$, which is slightly stronger on Callisto's trailing side. This band has not been previously noted on Callisto in either ground-based or NIMS spectra. Although a 3.78-$\micron$ band has been detected on Europa \citep{trumbo2017new}, the 3.75-$\micron$ band we have detected on Callisto is much narrower and weaker, and it is uncertain whether these two bands might share a common origin. 
	
\begin{figure}[h!]
	\centering
	\includegraphics[scale=0.85]{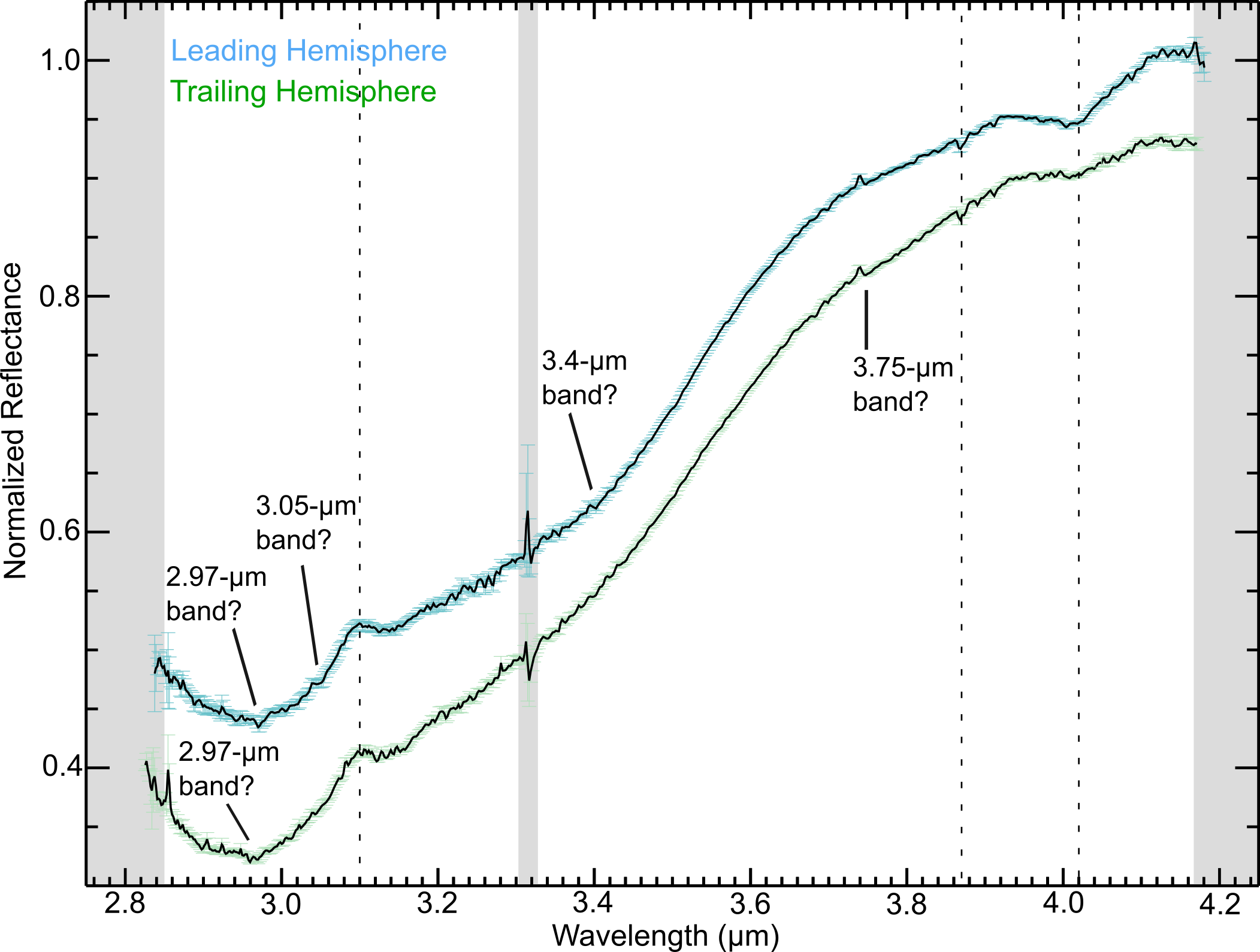}
	\caption{\textit{{Grand average spectra, and 1$\sigma$ uncertainties, of Callisto's leading and trailing hemisphere (blue and green error bars, respectively). These data are normalized to 1 at 4.12 $\micron$, and offset vertically for clarity. The dashed lines at 3.1, 3.87, and 4.02 $\micron$ highlight the central wavelengths of the H$_2$O ice Fresnel peak and the 3.9-$\micron$ and 4-$\micron$ bands, respectively. Subtle absorption bands hinted at in the grand average leading hemisphere spectrum data, centered near 2.97, 3.05, and 3.4 $\micron$,  and in the grand average trailing hemisphere spectrum, centered near 2.97 and 3.75 $\micron$, are indicated. Both spectra have been lightly smoothed using a 9 pixel-wide boxcar function. Wavelength regions where strong telluric bands are present are shown as gray-toned zones.}}}
\end{figure}

\newpage

\subsection{Results: 4-$\micron$ band parameter measurements}

\justify
In this appendix, we report the results of our \textit{F}-test analysis (Table A1).

\begin{table}[htbp]
	\caption {\textit{F}-test analysis of the longitudinal distribution of the 4-$\micron$ band.} 
	\hskip-2.8cm\begin{tabular}{llllllllllllll}
		\toprule
		\begin{tabular}[c]{@{}l@{}}\hspace{-1 cm}Band\\  \end{tabular} & \begin{tabular}[c]{@{}l@{}} \hspace{-1 cm}Band\\  \hspace{-1 cm}Measurement\\ \end{tabular}& \begin{tabular}[c]{@{}l@{}} \hspace{-1 cm}One Tailed \\  \hspace{-1 cm}\textit{F}-test Ratio \end{tabular}  & \begin{tabular}[c]{@{}l@{}} \hspace{-1 cm}Sample \\  \hspace{-1 cm}Size (n) \end{tabular} &
		\begin{tabular}[c]{@{}l@{}} \hspace{-1 cm}Mean Model \\ \hspace{-1 cm}Degree of \\ \hspace{-1 cm}Freedom (n-1) \end{tabular} & \begin{tabular}[c]{@{}l@{}} \hspace{-1 cm} Sinusoidal \\ \hspace{-1 cm}Model Degree of \\ \hspace{-1 cm}Freedom (n-3) \end{tabular} & \begin{tabular}[c]{@{}l@{}} \hspace{-1 cm}Probability (\textit{p}) \\  \end{tabular} & \begin{tabular}[c]{@{}l@{}} \hspace{-1 cm}Reject Null  \\  \hspace{-1 cm}Hypothesis? \end{tabular}\\
	  \midrule
		4-$\micron$ & Area &29.08 & 8 & 7 & 5 & 0.0001& Yes  \\
								 & Depth &24.21 & 8 & 7 & 5  & 0.0014& Yes \\						 
		\bottomrule
	\end{tabular}
\end{table}

\end{document}